\documentclass[%
reprint, 
 amsmath,amssymb,
prl,
]{revtex4-1}

\usepackage{graphicx}
\usepackage{dcolumn}
\usepackage{bm}
\usepackage{siunitx}

\begin{document}


\title{Coherent suppression of tensor frequency shifts through magnetic field rotation}

\author{R. Lange}
  \email{richard.lange@ptb.de}
\author{N. Huntemann}%
\author{C. Sanner}
\altaffiliation[Present address: ]{JILA, Boulder, CO 80309, USA}
\author{H. Shao}
\author{B. Lipphardt}
\author{Chr. Tamm}
\author{E. Peik}
\affiliation{%
 Physikalisch-Technische Bundesanstalt, Bundesallee 100, 38116 Braunschweig, Germany
}%

\date{\today}

\begin{abstract}
We introduce a scheme to coherently suppress second-rank tensor frequency shifts in atomic clocks, relying on the continuous rotation of an external magnetic field during the free atomic state evolution in a Ramsey sequence. The method retrieves the unperturbed frequency within a single interrogation cycle and is readily applicable to various atomic clock systems. For the frequency shift due to the electric quadrupole interaction, we experimentally demonstrate suppression by more than two orders of magnitude for the ${}^2S_{1/2} \to {}^2D_{3/2}$ transition of a single trapped ${}^{171}\text{Yb}^+$ ion. The scheme provides particular advantages in the case of the ${}^{171}\text{Yb}^+$ ${}^2S_{1/2} \to {}^2F_{7/2}$ electric octupole (E3) transition. For an improved estimate of the residual quadrupole shift for this transition, we measure the excited state electric quadrupole moments $\Theta({}^2D_{3/2}) = 1.95(1)~ea_0^2$ and $\Theta({}^2F_{7/2}) = -0.0297(5)~ea_0^2$ with $e$ the elementary charge and $a_0$ the Bohr radius, improving the measurement uncertainties by one order of magnitude. 
\end{abstract}

\pacs{Valid PACS appear here}
\maketitle


The ever-increasing level of control over neutral atoms and ions provides rapid progress in their wide range of applications from quantum computing to high-precision frequency measurements. For such measurements, optical atomic clocks are currently pushing the limits of relative uncertainty to the $10^{-19}$ range \cite{McGrew.2018,Brewer.2019,Ludlow.2015}. Precision and universality of atomic clocks are based on the concept of an unperturbed transition frequency, as observed under idealized conditions in the absence of systematic frequency shifts. To approach this situation, great care needs to be taken to control and reduce external perturbations. 

In many cases, clock data is corrected for frequency shifts in post-processing. Some perturbations can be suppressed in real time through operation at shift-free, so-called magic parameters. Prominent examples are the light shift in optical dipole traps \cite{Katori.2003,Ye.2008} or the mutual compensation of scalar Stark and relativistic Doppler shifts in Paul traps \cite{Dube.2014}. Other systematic shifts can be eliminated by averaging the clock output frequency over a set of discrete experimental parameters that are used in alternating cycles of clock operation. This includes probing different Zeeman components of the clock transition \cite{Bernard.1999,Dube.2005} and cycling between mutually orthogonal directions of an externally applied magnetic field defining the quantization axis\cite{Itano.2000}. Such schemes may contain nested feedback loops for the control of several parameters \cite{McGrew.2018,Brewer.2019,Sanner.2018}.

Coherent suppression is another form of shift elimination that relies on dynamic modification of experimental parameters during the atomic state evolution, providing the advantages of a fixed clock sequence and shift suppression within a single cycle. For optical clocks, different types of dynamic decoupling schemes have been investigated in which sublevels of the clock states are coherently coupled during the dark time of a Ramsey interrogation for cancellation of various shifts \cite{Kaewuam.2020,Shaniv.2019,Aharon.2019}. Further examples of coherent suppression can be found in Hyper-Ramsey spectroscopy dealing with light shifts \cite{Yudin.2010}, and entangled many-atom states designed to provide immunity from selected perturbations \cite{Roos.2006}. 

In the following, we focus on frequency shifts produced by orientation-dependent interactions described by traceless symmetric second-rank tensors. Common examples of such effects in atomic spectroscopy are the tensorial Stark effect, the quadrupole interaction with an electric field gradient, or the interaction between two magnetic dipoles. The symmetry properties of the second-degree spherical harmonics $Y_2^m$ that describe the orientation dependence of these interactions lead to vanishing frequency shifts for isotropic perturbations, such as spin-spin interactions in a liquid. They also allow for methods that eliminate the shift by averaging. A technique called magic angle spinning (MAS) is commonly used in nuclear magnetic resonance (NMR) spectroscopy of solids \cite{Andrew.2008}: The probe is rapidly spun around an axis oriented at the magic angle $\theta_m$ defined by $Y_2^0(\theta_m)=0$, corresponding to $\cos^2\theta_m=1/3$, with respect to the magnetic field. In the time average, the tensor shifts disappear and a significant narrowing of the NMR line can be observed in comparison to a stationary sample. A related method is used in trapped-ion optical clocks for the elimination of shifts resulting from tensor interactions with arbitrary orientation \cite{Itano.2000}: The clock frequency is recorded for three mutually orthogonal orientations of an external magnetic field so that the average value of the frequency is free from the tensor shift. As the result is obtained from three independent measurements, this scheme can be regarded as an incoherent averaging method.

The method presented in this letter can be seen as MAS over a single revolution or as a coherent and continuous version of the averaging over three mutually orthogonal orientations of the magnetic field. Rotating this system of orientations around its space diagonal, again leads to the angle $\theta_m$ between the axis of rotation and the magnetic field. Effectively, the magnetic field is rotated on the surface of a cone with the opening angle $\theta_m$. A similar magnetic field spinning technique has been applied to tune the magnetic dipole-dipole interaction in a Bose-Einstein condensate \cite{Giovanazzi.2002,Tang.2018}. 

\begin{figure*}[t]
\includegraphics[width = 0.9\textwidth]{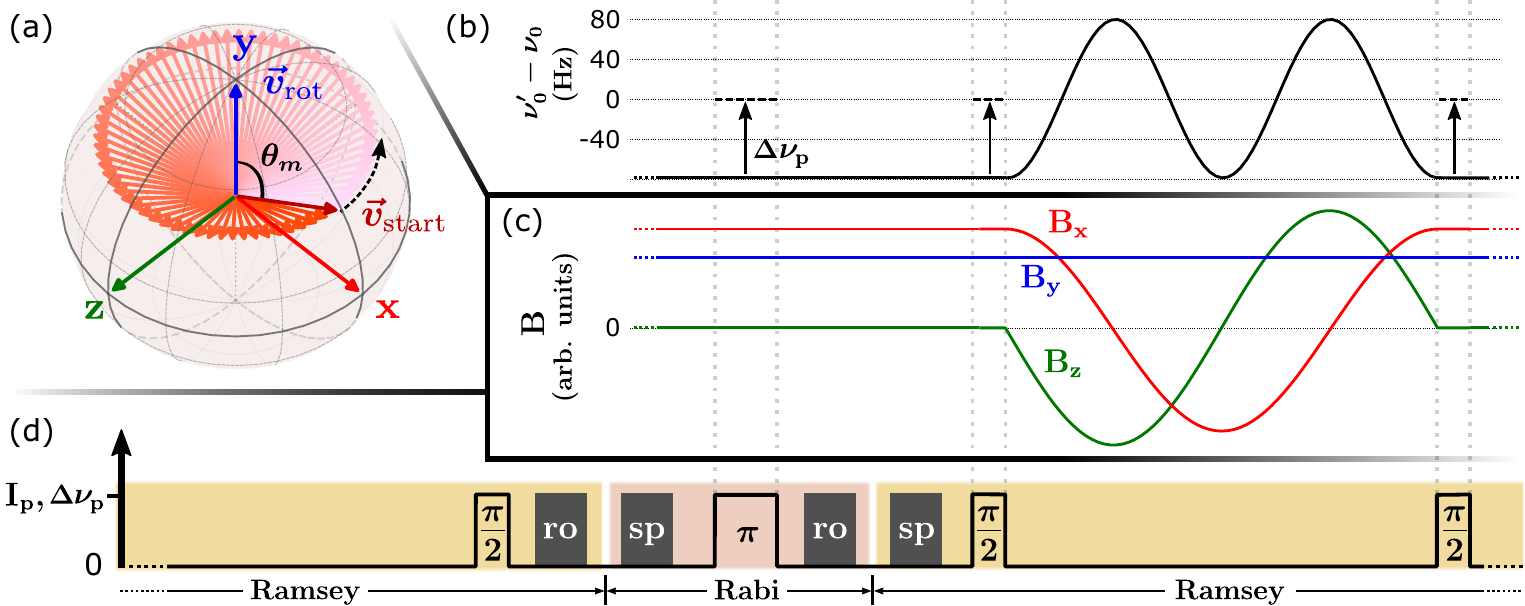}
\caption{Visualization of the coherent suppression sequence. (a) Starting from $\vec{v}_{\text{start}}$, the magnetic field is rotated around $\vec{v}_{\text{rot}}$ at constant cone angle $\theta_m$. (b) The instantaneous atomic resonance frequency $\nu_0'$, shifted by the electric field gradient, experiences an offset from the unperturbed transition frequency $\nu_0$ that averages to zero for the full rotation. The rotation is embedded in a Ramsey sequence probing $\nu_0$ in a feedback loop. During the $\pi/2$-pulses, an auxiliary frequency offset $\Delta\nu_{\text{p}}$ on the clock laser, derived from interleaved Rabi excitations, ensures resonant driving of the clock transition. (c) Temporal variation of the components of the magnetic field vectors. (d) Clock laser intensity I\textsubscript{p} and frequency deviation $\Delta\nu_{\text{p}}$ from $\nu_0$ for the Rabi-controlled Ramsey cycle. Readout (ro) and state preparation (sp) is performed in between interrogations.}
\label{graph:MagRot}
\end{figure*}

In our experiment, a single $2\pi$ rotation of the external magnetic field is performed during the free atomic state evolution between the two pulses of a Ramsey interrogation. Thereby the tensor shifts are averaged to zero and the clock laser frequency is coherently compared with the unperturbed atomic reference. It is assumed here that the tensor shift inducing field gradient does not change during the Ramsey interrogation. The time for one rotation of the magnetic field vector is chosen to be much longer than the inverse splitting frequency between Zeeman sublevels of the atomic states, so that atoms follow the magnetic field adiabatically, avoiding Majorana spin-flip transitions. In general, an explicitly time-dependent Hamiltonian can lead to a nonzero geometric phase \cite{Berry.1984,Suter.1987} in addition to the dynamic phase accumulated between the clock states. This phase, also known as Berry phase, scales with the enclosed solid angle along the magnetic field trajectory and can therefore easily be taken into account. Often, like in the case of the later considered ${}^{171}\text{Yb}^+$ clock states, the Berry phase is common-mode suppressed for the ground and excited states so that it does not contribute to the Ramsey signal \cite{Meyer.2009}.

While we demonstrate the coherent suppression for a simple single-ion optical atomic clock system, we see wide application in high-precision spectroscopy and other clock systems \cite{Campbell.2017,Huntemann.2016,Herschbach.2012,Arnold.2015,Golovizin.2019}. The method may be particularly beneficial for optical clocks based on ions in Coulomb crystals \cite{Herschbach.2012,Arnold.2015} . Here, the radio-frequency (RF) trapping field and large static electric field gradients typically cause tensor frequency shifts that exceed the resolvable linewidth resulting in inhomogeneous broadening, thus preventing long coherent interrogation. In contrast to dynamic decoupling schemes \cite{Shaniv.2019,Kaewuam.2020,Aharon.2019}, our method does not require any additional drive field that can also cause frequency shifts, but provides a straightforward solution for different optical clock systems seeking to mitigate tensor shifts. 

We experimentally demonstrate the method in a single-ion optical clock based on the  ${}^2S_{1/2} \to {}^2D_{3/2}$ electric quadrupole (E2) transition in ${}^{171}\text{Yb}^+$ \cite{Schneider.2005}. The coupling of the excited ${}^2D_{3/2}$ state to electric field gradients, externally applied or arising from spurious electric charges on the trap structure, makes this transition sensitive to the electric quadrupole shift (EQS). Using an endcap-trap design in our setup \cite{AbdelHafiz.2019}, a well-controlled electric field gradient can be generated through a static electric potential
\begin{align}
    \Phi_s (x',y',z') &= A [x'^2 (1+\epsilon) + y'^2 (1-\epsilon) - 2z'^2], 
    \label{eq:phi_static} \\
    A &= \frac{U}{\kappa}
     \label{eq:A}
\end{align}
by applying a voltage $U$ between inner and outer trap electrodes, with $\kappa$ a trap-dependent geometric factor, the coordinates $x', y', z'$ denoting the principal axes of the trap potential, and $\epsilon$ describing deviations from cylindrical symmetry of the trap.
The EQS can be expressed as \cite{Itano.2000}
\begin{align}
    \Delta \nu &= \frac{\alpha_k}{h} A \Theta [(3\cos^2 \theta_1 - 1) \nonumber \\ 
    &- \epsilon \sin^2 \theta_1 (\cos^2 \theta_2 - \sin^2 \theta_2) ]
    \label{eq:quadshift}
\end{align}
with $h$ the Planck constant, $\Theta$ the quadrupole moment and $\alpha_k$ containing the coupling terms of the Hamiltonian for transition $k$. In the case of ${}^{171}\text{Yb}^+$, $\alpha_{\text{E2}} = 1$ for the E2 transition and $\alpha_{\text{E3}} = 5/7$ for the electric octupole (E3) transition considered later. The angles $\theta_1$ and $\theta_2$ define the orientation of the principal axes of the field gradient relative to the magnetic field, with $\theta_1 = \sphericalangle (\hat{e}_{z'},\vec{B})$. The prefactors can be combined to an angle-independent frequency shift $\nu_{\text{quad}} = \alpha_k A \Theta / h$.

With three mutually orthogonal sets of coils, we define the unprimed coordinate system $xyz$ with $\hat{e}_z \parallel \hat{e}_{z'}$. The rotation axis $\vec{v}_{\text{rot}}$ of the magnetic field can be chosen arbitrarily. In our experimental demonstration, we perform rotations around the axes $x$, $y$ and $z$ with a magnetic field strength leading to a Zeeman splitting frequency $f_{Z} = 30$~kHz of the ${}^2D_{3/2}$ state. In the example shown in Fig. \ref{graph:MagRot}~(a), $\vec{v}_{\text{rot}} \parallel \hat{e}_y$. In that case, the initial magnetic field orientation is $\vec{v}_{\text{start}} = \sin(\theta_m) \hat{e}_x + \cos(\theta_m) \hat{e}_y$. Each vector $\vec{v}_{\text{path}}$ on the trajectory can now be addressed via
\begin{align}
    \vec{v}_{\text{path}}(\varphi) &= R(\vec{v}_{\text{rot}}, \varphi) \cdot \vec{v}_{\text{start}}
\end{align}
where $R(\vec{v}_{\text{rot}}, \varphi)$ is a rotation matrix around $\vec{v}_{\text{rot}}$ with rotation angle  $\varphi \in [0,2\pi]$. The instantaneous deviation of the atomic transition frequency $\nu_0'$ due to the EQS from its unperturbed value $\nu_0$ is visualized in Fig.~\ref{graph:MagRot}, averaging to zero for the full rotation. The rotation is embedded in a Ramsey sequence probing $\nu_0$ in a feedback loop. The frequency deviation at the initial magnetic field, present during the Ramsey pulses, is compensated by a clock laser offset $\Delta\nu$\textsubscript{p}. The size of the offset is determined with interleaved Rabi excitations in a second feedback loop \cite{Huntemann.2016}. The combined operation of the two feedback loops effectively cancels perturbations during the Ramsey pulses and allows us to realize the unperturbed transition frequency $\nu_0$.

In any experimental realization, the magnetic field does not instantaneously follow the applied coil currents due to delayed magnetization and eddy-current effects. To model the field response in our system, we apply a step function to the control input of the current driver for each set of coils and measure the magnetic field at the ion position after a variable time delay $t$ by means of the Zeeman splitting frequency $f_{Z}$. We describe the delayed field response by a linear combination of exponential functions
\begin{align}
    f_{Z}(t)  = (&f_{\text{start}}-f_{\text{end}}) \\
   (&A_1 e^{-(t-\tilde{t})/\tau_1} + A_2 e^{-(t-\tilde{t})/\tau_2}) + f_{\text{end}}, \nonumber
    \label{eq:decay}
\end{align}
with $f_{\text{start}}$ and $f_{\text{end}}$ the equilibrium Zeeman splittings before and after the step, and $A_1 + A_2 = 1$. For the three sets of coils, the inferred time constants $\tau_1$ and $\tau_2$ are approximately $0.2$ and $1$~\si{\milli\s}. Along the $x$-axis, an additional slow decay term with $\tau_3 = 16$~\si{\milli\s} is taken into account. Differences in the response times $\tilde{t}$  lead to distortions of the trajectory and are on the order of $0.1$~\si{\milli\s} for our coil pairs.

\begin{figure}[h]
\includegraphics[width = 0.48\textwidth]{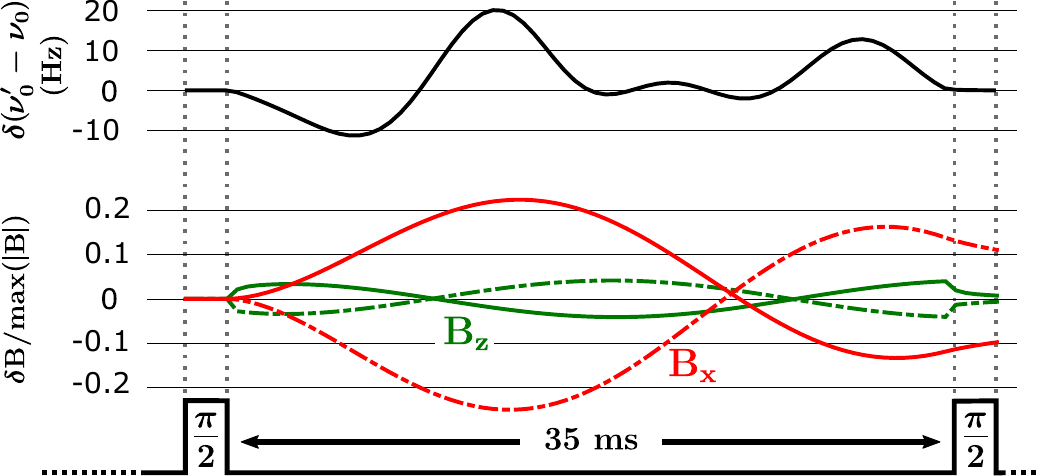}
\caption{Deviations of the instantaneous atomic resonance frequency (black line) and the magnetic field components (solid red and green line) due to the field delay from the intended trajectories displayed in Fig.~\ref{graph:MagRot}. Shown are simulations based on the measured step response of the magnetic field. The dashed lines describe the applied magnetic field compensations, correcting for the distorted trajectories.}
\label{graph:MagDev}
\end{figure}

Applying our field delay model, we test the distortion of the trajectories and the effect on $\nu_0'$ during the Ramsey dark time. Deviations of both the transition frequency and magnetic field are shown in Fig.~\ref{graph:MagDev}. In a second step, we design compensating field variations taking the delay into account such that the ideal rotation is retrieved. 

Our experimental investigation of the coherent suppression scheme is performed using the excited clock state ${}^2D_{3/2}$ that has a natural lifetime of 53~ms \cite{Yu.2000}. In practice this limits the available Ramsey dark times to a few tens of milliseconds, but the shift suppression due to the field rotation is not compromised because the Ramsey signal arises only from cycles in which the atom does not spontaneously decay to the ground state during the dark time. 

The EQS-free E2 transition frequency is initially determined by averaging the measured frequency along three orthogonal directions of the magnetic field \cite{Itano.2000}. An independent optical clock serves as a stable frequency reference. This measurement is performed once without and once with a large externally produced field gradient. Comparing the two measurements, we find a transition frequency difference of $0.02(4)$~Hz, consistent with zero, thus verifying the orthogonality of the employed magnetic field directions. All following coherent suppression experiments are performed with the external gradient that yields $\nu_{\text{quad}} = 78.7(2)$~Hz.

We first implement the coherent suppression scheme for a 35~ms Ramsey dark time without compensating for the magnetic field decay. For three measurements, each with $\vec{v}_{\text{rot}}$ along one of the $xyz$ coordinate axes, we find the frequency deviations $\Delta \nu^{\text{exp}}$ shown in Table~\ref{tab:rotFreqs}. With knowledge of strength and orientation of the EQS, we simulate the expected shift and find the frequency deviations $\Delta \nu^{\text{sim}}$. These values are corrected for a modified second-order Zeeman shift as the magnetic field strength varies during the distorted rotation by up to five percent.
We test our simulation by comparing it to the experimental values and calculating their difference as shown in Table~\ref{tab:rotFreqs}. Finally, compensating for the trajectory distortions, we find the frequency shifts $\Delta \nu^{\text{exp}}_{\text{comp}}$ from the unperturbed transition frequency. With the mean electric quadrupole shift $\nu_{\text{quad}} = 78.7(2)$~Hz and the largest deviation of $-0.3$~Hz, a suppression factor of 260 is derived. 

\begin{table}[htb]
    \centering
    \begin{tabular*}{0.48\textwidth}{c @{\extracolsep{\fill}} rrcr}
    $\vec{v}_{\text{rot}}$ & $\Delta \nu^{\text{exp}}$ & $\Delta \nu^{\text{sim}}$  &  $\Delta \nu^{\text{exp}} - \Delta \nu^{\text{sim}}$ & $\Delta \nu^{\text{exp}}_{\text{comp}}$ \\
    \hline 
    $\hat{e}_x$ & $-0.51 (4)$  & $-0.46 (10)$ & $-0.05 (11)$ & $-0.30 (8)$ \\
    $\hat{e}_y$ & $2.50 (4)$   & $2.67 (25)$ & $-0.17 (25)$  & $-0.05 (9)$ \\
    $\hat{e}_z$ & $8.07 (3)$   & $7.74 (30)$ & $\hphantom{-}0.33 (30)$  & $0.00 (7)$
    \end{tabular*}
    \caption{Deviations from the unperturbed transition frequency for rotation around three axes for a known externally produced electric field gradient. Without compensating the magnetic field decay, we find the experimental values $\Delta \nu^{\text{exp}}$. Simulations of the expected offsets $\Delta \nu^{\text{sim}}$ due to the magnetic field delay are compared with the experiment. The uncertainties of the simulations are derived from the variation of the decay times of the field delay model using a Monte-Carlo method. For the experimental results, statistical uncertainties are given. The coherent suppression scheme performed with the compensated magnetic field trajectory yields the residual frequency shifts $\Delta \nu^{\text{exp}}_{\text{comp}}$. All values are given in Hz.}
    \label{tab:rotFreqs}
\end{table}

The coherent suppression scheme, demonstrated for the ${}^{171}\text{Yb}^+$ E2 transition, can be readily applied to other clock types and species. One promising candidate for future application of this method is the ${}^2S_{1/2} \to {}^2F_{7/2}$ electric octupole (E3) \cite{Sanner.2019} transition in ${}^{171}\text{Yb}^+$. A natural excited state lifetime of years enables long Ramsey dark periods, limited only by the coherence time of the atom-laser interaction, permitting a slower rotation and consequently less deviation from the ideal trajectory than demonstrated on the E2 transition.
The excitation of the E3 transition requires high laser intensity due to the small oscillator strength, leading to a considerable light shift. Therefore, incoherent EQS cancellation based on sequential interrogation in three mutually orthogonal magnetic field orientations would result in large variations of the light shift, degrading clock performance. In contrast, the coherent suppression method allows for a free choice of the magnetic field orientation during the interrogation pulses. 
In order to estimate the smaller EQS on the E3 transition from the measured EQS on the E2 transition, the relative magnitude of the ${}^2D_{3/2}$ and ${}^2F_{7/2}$ quadrupole moments has to be known. So far, only one experimental investigation has been performed for each level \cite{Schneider.2005,Huntemann.2012} and the particularly small value of $\Theta({}^2F_{7/2})$ is not supported so far by theoretical studies \cite{Porsev.2012,Nandy.2014,Batra.2016,Guo.2020}. According to Eq.~\ref{eq:quadshift}, the experimental determination of the quadrupole moments requires knowledge of the applied electric field gradient. The geometric factor $\kappa$ in Eq.~\ref{eq:A} can be derived from the dependence of the ion secular frequencies on the applied voltage $U$ as presented in \cite{Bate.1992}. For the employed ion trap, we find $\kappa = -1.0417(8)~\si{\mm}^2$ and $\epsilon = 0.036(1)$. To infer the quadrupole moments, we measure the EQS for the E2 and E3 transition with a constant magnetic field orientation $\theta_1~=~1(1)^{\circ}$. In this configuration, the second term of Eq.~\ref{eq:quadshift} is negligible. 

\begin{figure}[h]
\includegraphics{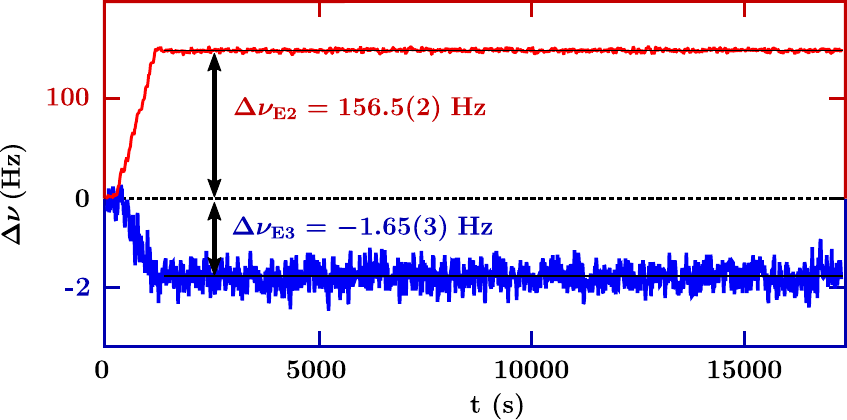}
\caption{Electric quadrupole shift $\Delta\nu$ due to an externally applied electric field gradient measured with interleaved operation of the clock, sequentially probing the E2 (red) and E3 (blue) transitions of a single trapped $^{171}$Yb$^+$ ion. Using an independent clock as the reference, the field gradient is slowly increased up to $A~=~57.6(3)~$V/mm$^2$, and the induced frequency offset is averaged for more than four hours. For both transitions, the unshifted frequency ($\Delta\nu=0$~Hz) is determined by preceding long-term measurements.}
\label{graph:QuadMomente_shift}
\end{figure}

We operate the experiment in an interleaved mode, sequentially probing the two clock transitions. The E2 transition is interrogated with \SI{40}{\milli\s} Rabi pulses. For the E3 transition, the light shift is cancelled using the Rabi-controlled Ramsey method with \SI{300}{\milli \s} dark periods \cite{Huntemann.2016}. We measure the transition frequencies for the selected magnetic field orientation with respect to an independent optical clock. As shown in Fig.~\ref{graph:QuadMomente_shift}, after a measurement period without applied field gradient, the voltage between the endcap electrodes is ramped up and kept at $-60.0(3)$~V, corresponding to $A~=~57.6(3)~$V/mm$^2$. From a measurement time of more than four hours, we find $\Delta \nu_{\text{E2}} = \SI{152.0(2)}{\Hz}$ and $\Delta \nu_{\text{E3}} = \SI{-1.65(3)}{\Hz}$ and calculate the quadrupole moments $\Theta({}^2D_{3/2}) = 1.95(1)~ea_0^2$ and $\Theta({}^2F_{7/2}) = -0.0297(5)~ea_0^2$ with $e$ the elementary charge and $a_0$ the Bohr radius. In comparison with previous investigations, the uncertainty is reduced for both values by about one order of magnitude. We respectively find a $1\sigma$- and a $2\sigma$-sigma agreement with previous measurements of $\Theta({}^2D_{3/2})$  \cite{Schneider.2005} and $\Theta({}^2F_{7/2})$ \cite{Huntemann.2012}. Theoretical predictions are within 10~\% for $\Theta({}^2D_{3/2})$  \cite{Itano.2000,Latha.2007,Nandy.2014,Batra.2016,Guo.2020}, but for $\Theta({}^2F_{7/2})$ they differ by at least a factor of two from the experimental value \cite{Porsev.2012,Nandy.2014,Batra.2016,Guo.2020}. 
This discrepancy appears to be related to the complex electronic structure of the ${}^2F_{7/2}$ state which makes calculations of the quadrupole moment challenging \cite{Porsev.2012}. 
Independent of the specific magnitude and orientation of the applied electric field gradient, the ratio of the measured frequency shifts $\Delta \nu_{\text{E2}}/\Delta \nu_{\text{E3}} = -92.1(1.7)$ yields the quadrupole moment ratio $\Theta({}^2D_{3/2})/\Theta({}^2F_{7/2}) = -65.8(1.2)$. 

In the employed ion trap, patch charges typically induce electric field gradients of less than $1~$V/mm$^2$, resulting in shifts on the order of $\nu_{\text{quad}}~=~1$~Hz on the E2 transition. With the demonstrated suppression factor of 260 and the hundredfold lower sensitivity of the E3 transition, we expect a residual shift of less than 0.04~mHz using the coherent suppression scheme. This corresponds to a negligible contribution of about $6\times 10^{-20}$ to the relative uncertainty of the E3 clock. 

Inspired by magic angle spinning in NMR spectroscopy, we have introduced a coherent suppression method for second-rank tensor interactions with possible application in a wide range of experiments on precision spectroscopy and quantum control. A dynamic magnetic field orientation provides a new parameter to obtain immunity to external perturbations in multi-pulse Ramsey schemes. While our implementation specifically aims for a simple sequence, we expect that this approach could be expanded to suppress additional shift effects or to enhance robustness by employing more complex field orientation trajectories and modulations of the field strength.

We thank T. Liebisch for a critical reading of the manuscript.
This work has been supported by the EMPIR projects 17FUN07 "Coulomb Crystals for Clocks" and 18SIB05 "Robust Optical Clocks for International Timescales". 
This project has received funding from the EMPIR programme co-financed
by the Participating States and from the European Union’s Horizon 2020
research and innovation programme.
This work has been supported by the Max-Planck-RIKEN-PTB-Center for Time, Constants and Fundamental Symmetries. C. Sanner thanks the Humboldt Foundation for support.

%


\begin{thebibliography}{37}%
\makeatletter
\providecommand \@ifxundefined [1]{%
 \@ifx{#1\undefined}
}%
\providecommand \@ifnum [1]{%
 \ifnum #1\expandafter \@firstoftwo
 \else \expandafter \@secondoftwo
 \fi
}%
\providecommand \@ifx [1]{%
 \ifx #1\expandafter \@firstoftwo
 \else \expandafter \@secondoftwo
 \fi
}%
\providecommand \natexlab [1]{#1}%
\providecommand \enquote  [1]{``#1''}%
\providecommand \bibnamefont  [1]{#1}%
\providecommand \bibfnamefont [1]{#1}%
\providecommand \citenamefont [1]{#1}%
\providecommand \href@noop [0]{\@secondoftwo}%
\providecommand \href [0]{\begingroup \@sanitize@url \@href}%
\providecommand \@href[1]{\@@startlink{#1}\@@href}%
\providecommand \@@href[1]{\endgroup#1\@@endlink}%
\providecommand \@sanitize@url [0]{\catcode `\\12\catcode `\$12\catcode
  `\&12\catcode `\#12\catcode `\^12\catcode `\_12\catcode `\%12\relax}%
\providecommand \@@startlink[1]{}%
\providecommand \@@endlink[0]{}%
\providecommand \url  [0]{\begingroup\@sanitize@url \@url }%
\providecommand \@url [1]{\endgroup\@href {#1}{\urlprefix }}%
\providecommand \urlprefix  [0]{URL }%
\providecommand \Eprint [0]{\href }%
\providecommand \doibase [0]{http://dx.doi.org/}%
\providecommand \selectlanguage [0]{\@gobble}%
\providecommand \bibinfo  [0]{\@secondoftwo}%
\providecommand \bibfield  [0]{\@secondoftwo}%
\providecommand \translation [1]{[#1]}%
\providecommand \BibitemOpen [0]{}%
\providecommand \bibitemStop [0]{}%
\providecommand \bibitemNoStop [0]{.\EOS\space}%
\providecommand \EOS [0]{\spacefactor3000\relax}%
\providecommand \BibitemShut  [1]{\csname bibitem#1\endcsname}%
\let\auto@bib@innerbib\@empty
\bibitem [{\citenamefont {McGrew}\ \emph {et~al.}(2018)\citenamefont {McGrew},
  \citenamefont {Zhang}, \citenamefont {Fasano}, \citenamefont {Sch{\"a}ffer},
  \citenamefont {Beloy}, \citenamefont {Nicolodi}, \citenamefont {Brown},
  \citenamefont {Hinkley}, \citenamefont {Milani}, \citenamefont {Schioppo},
  \citenamefont {Yoon},\ and\ \citenamefont {Ludlow}}]{McGrew.2018}%
  \BibitemOpen
  \bibfield  {author} {\bibinfo {author} {\bibfnamefont {W.~F.}\ \bibnamefont
  {McGrew}}, \bibinfo {author} {\bibfnamefont {X.}~\bibnamefont {Zhang}},
  \bibinfo {author} {\bibfnamefont {R.~J.}\ \bibnamefont {Fasano}}, \bibinfo
  {author} {\bibfnamefont {S.~A.}\ \bibnamefont {Sch{\"a}ffer}}, \bibinfo
  {author} {\bibfnamefont {K.}~\bibnamefont {Beloy}}, \bibinfo {author}
  {\bibfnamefont {D.}~\bibnamefont {Nicolodi}}, \bibinfo {author}
  {\bibfnamefont {R.~C.}\ \bibnamefont {Brown}}, \bibinfo {author}
  {\bibfnamefont {N.}~\bibnamefont {Hinkley}}, \bibinfo {author} {\bibfnamefont
  {G.}~\bibnamefont {Milani}}, \bibinfo {author} {\bibfnamefont
  {M.}~\bibnamefont {Schioppo}}, \bibinfo {author} {\bibfnamefont {T.~H.}\
  \bibnamefont {Yoon}}, \ and\ \bibinfo {author} {\bibfnamefont {A.~D.}\
  \bibnamefont {Ludlow}},\ }\href {\doibase 10.1038/s41586-018-0738-2}
  {\bibfield  {journal} {\bibinfo  {journal} {Nature}\ }\textbf {\bibinfo
  {volume} {564}},\ \bibinfo {pages} {87} (\bibinfo {year} {2018})}\BibitemShut
  {NoStop}%
\bibitem [{\citenamefont {Brewer}\ \emph {et~al.}(2019)\citenamefont {Brewer},
  \citenamefont {Chen}, \citenamefont {Hankin}, \citenamefont {Clements},
  \citenamefont {Chou}, \citenamefont {Wineland}, \citenamefont {Hume},\ and\
  \citenamefont {Leibrandt}}]{Brewer.2019}%
  \BibitemOpen
  \bibfield  {author} {\bibinfo {author} {\bibfnamefont {S.~M.}\ \bibnamefont
  {Brewer}}, \bibinfo {author} {\bibfnamefont {J.-S.}\ \bibnamefont {Chen}},
  \bibinfo {author} {\bibfnamefont {A.~M.}\ \bibnamefont {Hankin}}, \bibinfo
  {author} {\bibfnamefont {E.~R.}\ \bibnamefont {Clements}}, \bibinfo {author}
  {\bibfnamefont {C.~W.}\ \bibnamefont {Chou}}, \bibinfo {author}
  {\bibfnamefont {D.~J.}\ \bibnamefont {Wineland}}, \bibinfo {author}
  {\bibfnamefont {D.~B.}\ \bibnamefont {Hume}}, \ and\ \bibinfo {author}
  {\bibfnamefont {D.~R.}\ \bibnamefont {Leibrandt}},\ }\href {\doibase
  10.1103/PhysRevLett.123.033201} {\bibfield  {journal} {\bibinfo  {journal}
  {Phys. Rev. Lett.}\ }\textbf {\bibinfo {volume} {123}},\ \bibinfo {pages}
  {033201} (\bibinfo {year} {2019})}\BibitemShut {NoStop}%
\bibitem [{\citenamefont {Ludlow}\ \emph {et~al.}(2015)\citenamefont {Ludlow},
  \citenamefont {Boyd}, \citenamefont {Ye}, \citenamefont {Peik},\ and\
  \citenamefont {Schmidt}}]{Ludlow.2015}%
  \BibitemOpen
  \bibfield  {author} {\bibinfo {author} {\bibfnamefont {A.~D.}\ \bibnamefont
  {Ludlow}}, \bibinfo {author} {\bibfnamefont {M.~M.}\ \bibnamefont {Boyd}},
  \bibinfo {author} {\bibfnamefont {J.}~\bibnamefont {Ye}}, \bibinfo {author}
  {\bibfnamefont {E.}~\bibnamefont {Peik}}, \ and\ \bibinfo {author}
  {\bibfnamefont {P.~O.}\ \bibnamefont {Schmidt}},\ }\href {\doibase
  10.1103/RevModPhys.87.637} {\bibfield  {journal} {\bibinfo  {journal} {Rev.
  Mod. Phys.}\ }\textbf {\bibinfo {volume} {87}},\ \bibinfo {pages} {637}
  (\bibinfo {year} {2015})}\BibitemShut {NoStop}%
\bibitem [{\citenamefont {Katori}\ \emph {et~al.}(2003)\citenamefont {Katori},
  \citenamefont {Takamoto}, \citenamefont {Pal'chikov},\ and\ \citenamefont
  {Ovsiannikov}}]{Katori.2003}%
  \BibitemOpen
  \bibfield  {author} {\bibinfo {author} {\bibfnamefont {H.}~\bibnamefont
  {Katori}}, \bibinfo {author} {\bibfnamefont {M.}~\bibnamefont {Takamoto}},
  \bibinfo {author} {\bibfnamefont {V.~G.}\ \bibnamefont {Pal'chikov}}, \ and\
  \bibinfo {author} {\bibfnamefont {V.~D.}\ \bibnamefont {Ovsiannikov}},\
  }\href {\doibase 10.1103/PhysRevLett.91.173005} {\bibfield  {journal}
  {\bibinfo  {journal} {Phys. Rev. Lett.}\ }\textbf {\bibinfo {volume} {91}},\
  \bibinfo {pages} {173005} (\bibinfo {year} {2003})}\BibitemShut {NoStop}%
\bibitem [{\citenamefont {Ye}\ \emph {et~al.}(2008)\citenamefont {Ye},
  \citenamefont {Kimble},\ and\ \citenamefont {Katori}}]{Ye.2008}%
  \BibitemOpen
  \bibfield  {author} {\bibinfo {author} {\bibfnamefont {J.}~\bibnamefont
  {Ye}}, \bibinfo {author} {\bibfnamefont {H.~J.}\ \bibnamefont {Kimble}}, \
  and\ \bibinfo {author} {\bibfnamefont {H.}~\bibnamefont {Katori}},\ }\href
  {\doibase 10.1126/science.1148259} {\bibfield  {journal} {\bibinfo  {journal}
  {Science}\ }\textbf {\bibinfo {volume} {320}},\ \bibinfo {pages} {1734}
  (\bibinfo {year} {2008})}\BibitemShut {NoStop}%
\bibitem [{\citenamefont {Dub{\'e}}\ \emph {et~al.}(2014)\citenamefont
  {Dub{\'e}}, \citenamefont {Madej}, \citenamefont {Tibbo},\ and\ \citenamefont
  {Bernard}}]{Dube.2014}%
  \BibitemOpen
  \bibfield  {author} {\bibinfo {author} {\bibfnamefont {P.}~\bibnamefont
  {Dub{\'e}}}, \bibinfo {author} {\bibfnamefont {A.~A.}\ \bibnamefont {Madej}},
  \bibinfo {author} {\bibfnamefont {M.}~\bibnamefont {Tibbo}}, \ and\ \bibinfo
  {author} {\bibfnamefont {J.~E.}\ \bibnamefont {Bernard}},\ }\href {\doibase
  10.1103/PhysRevLett.112.173002} {\bibfield  {journal} {\bibinfo  {journal}
  {Phys. Rev. Lett.}\ }\textbf {\bibinfo {volume} {112}},\ \bibinfo {pages}
  {173002} (\bibinfo {year} {2014})}\BibitemShut {NoStop}%
\bibitem [{\citenamefont {Bernard}\ \emph {et~al.}(1999)\citenamefont
  {Bernard}, \citenamefont {Madej}, \citenamefont {Marmet}, \citenamefont
  {Whitford}, \citenamefont {Siemsen},\ and\ \citenamefont
  {Cundy}}]{Bernard.1999}%
  \BibitemOpen
  \bibfield  {author} {\bibinfo {author} {\bibfnamefont {J.~E.}\ \bibnamefont
  {Bernard}}, \bibinfo {author} {\bibfnamefont {A.~A.}\ \bibnamefont {Madej}},
  \bibinfo {author} {\bibfnamefont {L.}~\bibnamefont {Marmet}}, \bibinfo
  {author} {\bibfnamefont {B.~G.}\ \bibnamefont {Whitford}}, \bibinfo {author}
  {\bibfnamefont {K.~J.}\ \bibnamefont {Siemsen}}, \ and\ \bibinfo {author}
  {\bibfnamefont {S.}~\bibnamefont {Cundy}},\ }\href {\doibase
  10.1103/PhysRevLett.82.3228} {\bibfield  {journal} {\bibinfo  {journal}
  {Phys. Rev. Lett.}\ }\textbf {\bibinfo {volume} {82}},\ \bibinfo {pages}
  {3228} (\bibinfo {year} {1999})}\BibitemShut {NoStop}%
\bibitem [{\citenamefont {Dub{\'e}}\ \emph {et~al.}(2005)\citenamefont
  {Dub{\'e}}, \citenamefont {Madej}, \citenamefont {Bernard}, \citenamefont
  {Marmet}, \citenamefont {Boulanger},\ and\ \citenamefont
  {Cundy}}]{Dube.2005}%
  \BibitemOpen
  \bibfield  {author} {\bibinfo {author} {\bibfnamefont {P.}~\bibnamefont
  {Dub{\'e}}}, \bibinfo {author} {\bibfnamefont {A.~A.}\ \bibnamefont {Madej}},
  \bibinfo {author} {\bibfnamefont {J.~E.}\ \bibnamefont {Bernard}}, \bibinfo
  {author} {\bibfnamefont {L.}~\bibnamefont {Marmet}}, \bibinfo {author}
  {\bibfnamefont {J.-S.}\ \bibnamefont {Boulanger}}, \ and\ \bibinfo {author}
  {\bibfnamefont {S.}~\bibnamefont {Cundy}},\ }\href {\doibase
  10.1103/PhysRevLett.95.033001} {\bibfield  {journal} {\bibinfo  {journal}
  {Phys. Rev. Lett.}\ }\textbf {\bibinfo {volume} {95}},\ \bibinfo {pages}
  {033001} (\bibinfo {year} {2005})}\BibitemShut {NoStop}%
\bibitem [{\citenamefont {Itano}(2000)}]{Itano.2000}%
  \BibitemOpen
  \bibfield  {author} {\bibinfo {author} {\bibfnamefont {W.~M.}\ \bibnamefont
  {Itano}},\ }\href {\doibase 10.6028/jres.105.065.} {\bibfield  {journal}
  {\bibinfo  {journal} {Journal of research of the National Institute of
  Standards and Technology}\ }\textbf {\bibinfo {volume} {105}},\ \bibinfo
  {pages} {829} (\bibinfo {year} {2000})}\BibitemShut {NoStop}%
\bibitem [{\citenamefont {Sanner}\ \emph {et~al.}(2018)\citenamefont {Sanner},
  \citenamefont {Huntemann}, \citenamefont {Lange}, \citenamefont {Tamm},\ and\
  \citenamefont {Peik}}]{Sanner.2018}%
  \BibitemOpen
  \bibfield  {author} {\bibinfo {author} {\bibfnamefont {C.}~\bibnamefont
  {Sanner}}, \bibinfo {author} {\bibfnamefont {N.}~\bibnamefont {Huntemann}},
  \bibinfo {author} {\bibfnamefont {R.}~\bibnamefont {Lange}}, \bibinfo
  {author} {\bibfnamefont {{\relax Chr}.}~\bibnamefont {Tamm}}, \ and\ \bibinfo
  {author} {\bibfnamefont {E.}~\bibnamefont {Peik}},\ }\href {\doibase
  10.1103/PhysRevLett.120.053602} {\bibfield  {journal} {\bibinfo  {journal}
  {Phys. Rev. Lett.}\ }\textbf {\bibinfo {volume} {120}},\ \bibinfo {pages}
  {053602} (\bibinfo {year} {2018})}\BibitemShut {NoStop}%
\bibitem [{\citenamefont {Kaewuam}\ \emph {et~al.}(2020)\citenamefont
  {Kaewuam}, \citenamefont {Tan}, \citenamefont {Arnold}, \citenamefont
  {Chanu}, \citenamefont {Zhang},\ and\ \citenamefont
  {Barrett}}]{Kaewuam.2020}%
  \BibitemOpen
  \bibfield  {author} {\bibinfo {author} {\bibfnamefont {R.}~\bibnamefont
  {Kaewuam}}, \bibinfo {author} {\bibfnamefont {T.~R.}\ \bibnamefont {Tan}},
  \bibinfo {author} {\bibfnamefont {K.~J.}\ \bibnamefont {Arnold}}, \bibinfo
  {author} {\bibfnamefont {S.~R.}\ \bibnamefont {Chanu}}, \bibinfo {author}
  {\bibfnamefont {Z.}~\bibnamefont {Zhang}}, \ and\ \bibinfo {author}
  {\bibfnamefont {M.~D.}\ \bibnamefont {Barrett}},\ }\href {\doibase
  10.1103/PhysRevLett.124.083202} {\bibfield  {journal} {\bibinfo  {journal}
  {Phys. Rev. Lett.}\ }\textbf {\bibinfo {volume} {124}},\ \bibinfo {pages}
  {083202} (\bibinfo {year} {2020})}\BibitemShut {NoStop}%
\bibitem [{\citenamefont {Shaniv}\ \emph {et~al.}(2019)\citenamefont {Shaniv},
  \citenamefont {Akerman}, \citenamefont {Manovitz}, \citenamefont {Shapira},\
  and\ \citenamefont {Ozeri}}]{Shaniv.2019}%
  \BibitemOpen
  \bibfield  {author} {\bibinfo {author} {\bibfnamefont {R.}~\bibnamefont
  {Shaniv}}, \bibinfo {author} {\bibfnamefont {N.}~\bibnamefont {Akerman}},
  \bibinfo {author} {\bibfnamefont {T.}~\bibnamefont {Manovitz}}, \bibinfo
  {author} {\bibfnamefont {Y.}~\bibnamefont {Shapira}}, \ and\ \bibinfo
  {author} {\bibfnamefont {R.}~\bibnamefont {Ozeri}},\ }\href {\doibase
  10.1103/PhysRevLett.122.223204} {\bibfield  {journal} {\bibinfo  {journal}
  {Phys. Rev. Lett.}\ }\textbf {\bibinfo {volume} {122}},\ \bibinfo {pages}
  {223204} (\bibinfo {year} {2019})}\BibitemShut {NoStop}%
\bibitem [{\citenamefont {Aharon}\ \emph {et~al.}(2019)\citenamefont {Aharon},
  \citenamefont {Spethmann}, \citenamefont {Leroux}, \citenamefont {Schmidt},\
  and\ \citenamefont {Retzker}}]{Aharon.2019}%
  \BibitemOpen
  \bibfield  {author} {\bibinfo {author} {\bibfnamefont {N.}~\bibnamefont
  {Aharon}}, \bibinfo {author} {\bibfnamefont {N.}~\bibnamefont {Spethmann}},
  \bibinfo {author} {\bibfnamefont {I.~D.}\ \bibnamefont {Leroux}}, \bibinfo
  {author} {\bibfnamefont {P.~O.}\ \bibnamefont {Schmidt}}, \ and\ \bibinfo
  {author} {\bibfnamefont {A.}~\bibnamefont {Retzker}},\ }\href {\doibase
  10.1088/1367-2630/ab3871} {\bibfield  {journal} {\bibinfo  {journal} {New
  Journal of Physics}\ }\textbf {\bibinfo {volume} {21}},\ \bibinfo {pages}
  {083040} (\bibinfo {year} {2019})}\BibitemShut {NoStop}%
\bibitem [{\citenamefont {Yudin}\ \emph {et~al.}(2010)\citenamefont {Yudin},
  \citenamefont {Taichenachev}, \citenamefont {Oates}, \citenamefont {Barber},
  \citenamefont {Lemke}, \citenamefont {Ludlow}, \citenamefont {Sterr},
  \citenamefont {Lisdat},\ and\ \citenamefont {Riehle}}]{Yudin.2010}%
  \BibitemOpen
  \bibfield  {author} {\bibinfo {author} {\bibfnamefont {V.~I.}\ \bibnamefont
  {Yudin}}, \bibinfo {author} {\bibfnamefont {A.~V.}\ \bibnamefont
  {Taichenachev}}, \bibinfo {author} {\bibfnamefont {C.~W.}\ \bibnamefont
  {Oates}}, \bibinfo {author} {\bibfnamefont {Z.~W.}\ \bibnamefont {Barber}},
  \bibinfo {author} {\bibfnamefont {N.~D.}\ \bibnamefont {Lemke}}, \bibinfo
  {author} {\bibfnamefont {A.~D.}\ \bibnamefont {Ludlow}}, \bibinfo {author}
  {\bibfnamefont {U.}~\bibnamefont {Sterr}}, \bibinfo {author} {\bibfnamefont
  {C.}~\bibnamefont {Lisdat}}, \ and\ \bibinfo {author} {\bibfnamefont
  {F.}~\bibnamefont {Riehle}},\ }\href {\doibase 10.1103/PhysRevA.82.011804}
  {\bibfield  {journal} {\bibinfo  {journal} {Phys. Rev. A}\ }\textbf {\bibinfo
  {volume} {82}},\ \bibinfo {pages} {011804(R)} (\bibinfo {year}
  {2010})}\BibitemShut {NoStop}%
\bibitem [{\citenamefont {Roos}\ \emph {et~al.}(2006)\citenamefont {Roos},
  \citenamefont {Chwalla}, \citenamefont {Kim}, \citenamefont {Riebe},\ and\
  \citenamefont {Blatt}}]{Roos.2006}%
  \BibitemOpen
  \bibfield  {author} {\bibinfo {author} {\bibfnamefont {C.~F.}\ \bibnamefont
  {Roos}}, \bibinfo {author} {\bibfnamefont {M.}~\bibnamefont {Chwalla}},
  \bibinfo {author} {\bibfnamefont {K.}~\bibnamefont {Kim}}, \bibinfo {author}
  {\bibfnamefont {M.}~\bibnamefont {Riebe}}, \ and\ \bibinfo {author}
  {\bibfnamefont {R.}~\bibnamefont {Blatt}},\ }\href {\doibase
  10.1038/nature05101} {\bibfield  {journal} {\bibinfo  {journal} {Nature}\
  }\textbf {\bibinfo {volume} {443}},\ \bibinfo {pages} {316} (\bibinfo {year}
  {2006})}\BibitemShut {NoStop}%
\bibitem [{\citenamefont {Andrew}(2008)}]{Andrew.2008}%
  \BibitemOpen
  \bibfield  {author} {\bibinfo {author} {\bibfnamefont {E.~R.}\ \bibnamefont
  {Andrew}},\ }\href {\doibase 10.1080/01442358109353320} {\bibfield  {journal}
  {\bibinfo  {journal} {International Reviews in Physical Chemistry}\ }\textbf
  {\bibinfo {volume} {1}},\ \bibinfo {pages} {195} (\bibinfo {year}
  {2008})}\BibitemShut {NoStop}%
\bibitem [{\citenamefont {Giovanazzi}\ \emph {et~al.}(2002)\citenamefont
  {Giovanazzi}, \citenamefont {G{\"o}rlitz},\ and\ \citenamefont
  {Pfau}}]{Giovanazzi.2002}%
  \BibitemOpen
  \bibfield  {author} {\bibinfo {author} {\bibfnamefont {S.}~\bibnamefont
  {Giovanazzi}}, \bibinfo {author} {\bibfnamefont {A.}~\bibnamefont
  {G{\"o}rlitz}}, \ and\ \bibinfo {author} {\bibfnamefont {T.}~\bibnamefont
  {Pfau}},\ }\href {\doibase 10.1103/PhysRevLett.89.130401} {\bibfield
  {journal} {\bibinfo  {journal} {Phys. Rev. Lett.}\ }\textbf {\bibinfo
  {volume} {89}},\ \bibinfo {pages} {130401} (\bibinfo {year}
  {2002})}\BibitemShut {NoStop}%
\bibitem [{\citenamefont {Tang}\ \emph {et~al.}(2018)\citenamefont {Tang},
  \citenamefont {Kao}, \citenamefont {Li},\ and\ \citenamefont
  {Lev}}]{Tang.2018}%
  \BibitemOpen
  \bibfield  {author} {\bibinfo {author} {\bibfnamefont {Y.}~\bibnamefont
  {Tang}}, \bibinfo {author} {\bibfnamefont {W.}~\bibnamefont {Kao}}, \bibinfo
  {author} {\bibfnamefont {K.-Y.}\ \bibnamefont {Li}}, \ and\ \bibinfo {author}
  {\bibfnamefont {B.~L.}\ \bibnamefont {Lev}},\ }\href {\doibase
  10.1103/PhysRevLett.120.230401} {\bibfield  {journal} {\bibinfo  {journal}
  {Phys. Rev. Lett.}\ }\textbf {\bibinfo {volume} {120}},\ \bibinfo {pages}
  {230401} (\bibinfo {year} {2018})}\BibitemShut {NoStop}%
\bibitem [{\citenamefont {Berry}(1984)}]{Berry.1984}%
  \BibitemOpen
  \bibfield  {author} {\bibinfo {author} {\bibfnamefont {M.~V.}\ \bibnamefont
  {Berry}},\ }\href {\doibase 10.1098/rspa.1984.0023} {\bibfield  {journal}
  {\bibinfo  {journal} {Proc. R. Soc. Lond. A}\ }\textbf {\bibinfo {volume}
  {392}},\ \bibinfo {pages} {45} (\bibinfo {year} {1984})}\BibitemShut
  {NoStop}%
\bibitem [{\citenamefont {Suter}\ \emph {et~al.}(1987)\citenamefont {Suter},
  \citenamefont {Chingas}, \citenamefont {Harris},\ and\ \citenamefont
  {Pines}}]{Suter.1987}%
  \BibitemOpen
  \bibfield  {author} {\bibinfo {author} {\bibfnamefont {D.}~\bibnamefont
  {Suter}}, \bibinfo {author} {\bibfnamefont {G.~C.}\ \bibnamefont {Chingas}},
  \bibinfo {author} {\bibfnamefont {R.~A.}\ \bibnamefont {Harris}}, \ and\
  \bibinfo {author} {\bibfnamefont {A.}~\bibnamefont {Pines}},\ }\href
  {\doibase 10.1080/00268978700101831} {\bibfield  {journal} {\bibinfo
  {journal} {Molecular Physics}\ }\textbf {\bibinfo {volume} {61}},\ \bibinfo
  {pages} {1327} (\bibinfo {year} {1987})}\BibitemShut {NoStop}%
\bibitem [{\citenamefont {Meyer}\ \emph {et~al.}(2009)\citenamefont {Meyer},
  \citenamefont {Leanhardt}, \citenamefont {Cornell},\ and\ \citenamefont
  {Bohn}}]{Meyer.2009}%
  \BibitemOpen
  \bibfield  {author} {\bibinfo {author} {\bibfnamefont {E.~R.}\ \bibnamefont
  {Meyer}}, \bibinfo {author} {\bibfnamefont {A.~E.}\ \bibnamefont
  {Leanhardt}}, \bibinfo {author} {\bibfnamefont {E.~A.}\ \bibnamefont
  {Cornell}}, \ and\ \bibinfo {author} {\bibfnamefont {J.~L.}\ \bibnamefont
  {Bohn}},\ }\href {\doibase 10.1103/PhysRevA.80.062110} {\bibfield  {journal}
  {\bibinfo  {journal} {Phys. Rev. A}\ }\textbf {\bibinfo {volume} {80}},\
  \bibinfo {pages} {062110} (\bibinfo {year} {2009})}\BibitemShut {NoStop}%
\bibitem [{\citenamefont {Campbell}\ \emph {et~al.}(2017)\citenamefont
  {Campbell}, \citenamefont {Hutson}, \citenamefont {Marti}, \citenamefont
  {Goban}, \citenamefont {{Darkwah Oppong}}, \citenamefont {McNally},
  \citenamefont {Sonderhouse}, \citenamefont {Robinson}, \citenamefont {Zhang},
  \citenamefont {Bloom},\ and\ \citenamefont {Ye}}]{Campbell.2017}%
  \BibitemOpen
  \bibfield  {author} {\bibinfo {author} {\bibfnamefont {S.~L.}\ \bibnamefont
  {Campbell}}, \bibinfo {author} {\bibfnamefont {R.~B.}\ \bibnamefont
  {Hutson}}, \bibinfo {author} {\bibfnamefont {G.~E.}\ \bibnamefont {Marti}},
  \bibinfo {author} {\bibfnamefont {A.}~\bibnamefont {Goban}}, \bibinfo
  {author} {\bibfnamefont {N.}~\bibnamefont {{Darkwah Oppong}}}, \bibinfo
  {author} {\bibfnamefont {R.~L.}\ \bibnamefont {McNally}}, \bibinfo {author}
  {\bibfnamefont {L.}~\bibnamefont {Sonderhouse}}, \bibinfo {author}
  {\bibfnamefont {J.~M.}\ \bibnamefont {Robinson}}, \bibinfo {author}
  {\bibfnamefont {W.}~\bibnamefont {Zhang}}, \bibinfo {author} {\bibfnamefont
  {B.~J.}\ \bibnamefont {Bloom}}, \ and\ \bibinfo {author} {\bibfnamefont
  {J.}~\bibnamefont {Ye}},\ }\href {\doibase 10.1126/science.aam5538}
  {\bibfield  {journal} {\bibinfo  {journal} {Science}\ }\textbf {\bibinfo
  {volume} {358}},\ \bibinfo {pages} {90} (\bibinfo {year} {2017})}\BibitemShut
  {NoStop}%
\bibitem [{\citenamefont {Huntemann}\ \emph {et~al.}(2016)\citenamefont
  {Huntemann}, \citenamefont {Sanner}, \citenamefont {Lipphardt}, \citenamefont
  {Tamm},\ and\ \citenamefont {Peik}}]{Huntemann.2016}%
  \BibitemOpen
  \bibfield  {author} {\bibinfo {author} {\bibfnamefont {N.}~\bibnamefont
  {Huntemann}}, \bibinfo {author} {\bibfnamefont {C.}~\bibnamefont {Sanner}},
  \bibinfo {author} {\bibfnamefont {B.}~\bibnamefont {Lipphardt}}, \bibinfo
  {author} {\bibfnamefont {{\relax Chr}.}~\bibnamefont {Tamm}}, \ and\ \bibinfo
  {author} {\bibfnamefont {E.}~\bibnamefont {Peik}},\ }\href {\doibase
  10.1103/PhysRevLett.116.063001} {\bibfield  {journal} {\bibinfo  {journal}
  {Phys. Rev. Lett.}\ }\textbf {\bibinfo {volume} {116}},\ \bibinfo {pages}
  {063001} (\bibinfo {year} {2016})}\BibitemShut {NoStop}%
\bibitem [{\citenamefont {Herschbach}\ \emph {et~al.}(2012)\citenamefont
  {Herschbach}, \citenamefont {Pyka}, \citenamefont {Keller},\ and\
  \citenamefont {Mehlst{\"a}ubler}}]{Herschbach.2012}%
  \BibitemOpen
  \bibfield  {author} {\bibinfo {author} {\bibfnamefont {N.}~\bibnamefont
  {Herschbach}}, \bibinfo {author} {\bibfnamefont {K.}~\bibnamefont {Pyka}},
  \bibinfo {author} {\bibfnamefont {J.}~\bibnamefont {Keller}}, \ and\ \bibinfo
  {author} {\bibfnamefont {T.~E.}\ \bibnamefont {Mehlst{\"a}ubler}},\ }\href
  {\doibase 10.1007/s00340-011-4790-y} {\bibfield  {journal} {\bibinfo
  {journal} {Appl. Phys. B}\ }\textbf {\bibinfo {volume} {107}},\ \bibinfo
  {pages} {891} (\bibinfo {year} {2012})}\BibitemShut {NoStop}%
\bibitem [{\citenamefont {Arnold}\ \emph {et~al.}(2015)\citenamefont {Arnold},
  \citenamefont {Hajiyev}, \citenamefont {Paez}, \citenamefont {Lee},
  \citenamefont {Barrett},\ and\ \citenamefont {Bollinger}}]{Arnold.2015}%
  \BibitemOpen
  \bibfield  {author} {\bibinfo {author} {\bibfnamefont {K.}~\bibnamefont
  {Arnold}}, \bibinfo {author} {\bibfnamefont {E.}~\bibnamefont {Hajiyev}},
  \bibinfo {author} {\bibfnamefont {E.}~\bibnamefont {Paez}}, \bibinfo {author}
  {\bibfnamefont {C.~H.}\ \bibnamefont {Lee}}, \bibinfo {author} {\bibfnamefont
  {M.~D.}\ \bibnamefont {Barrett}}, \ and\ \bibinfo {author} {\bibfnamefont
  {J.}~\bibnamefont {Bollinger}},\ }\href {\doibase 10.1103/PhysRevA.92.032108}
  {\bibfield  {journal} {\bibinfo  {journal} {Phys. Rev. A}\ }\textbf {\bibinfo
  {volume} {92}},\ \bibinfo {pages} {032108} (\bibinfo {year}
  {2015})}\BibitemShut {NoStop}%
\bibitem [{\citenamefont {Golovizin}\ \emph {et~al.}(2019)\citenamefont
  {Golovizin}, \citenamefont {Fedorova}, \citenamefont {Tregubov},
  \citenamefont {Sukachev}, \citenamefont {Khabarova}, \citenamefont
  {Sorokin},\ and\ \citenamefont {Kolachevsky}}]{Golovizin.2019}%
  \BibitemOpen
  \bibfield  {author} {\bibinfo {author} {\bibfnamefont {A.}~\bibnamefont
  {Golovizin}}, \bibinfo {author} {\bibfnamefont {E.}~\bibnamefont {Fedorova}},
  \bibinfo {author} {\bibfnamefont {D.}~\bibnamefont {Tregubov}}, \bibinfo
  {author} {\bibfnamefont {D.}~\bibnamefont {Sukachev}}, \bibinfo {author}
  {\bibfnamefont {K.}~\bibnamefont {Khabarova}}, \bibinfo {author}
  {\bibfnamefont {V.}~\bibnamefont {Sorokin}}, \ and\ \bibinfo {author}
  {\bibfnamefont {N.}~\bibnamefont {Kolachevsky}},\ }\href {\doibase
  10.1038/s41467-019-09706-9} {\bibfield  {journal} {\bibinfo  {journal}
  {Nature Comm.}\ }\textbf {\bibinfo {volume} {10}},\ \bibinfo {pages} {1724}
  (\bibinfo {year} {2019})}\BibitemShut {NoStop}%
\bibitem [{\citenamefont {Schneider}\ \emph {et~al.}(2005)\citenamefont
  {Schneider}, \citenamefont {Peik},\ and\ \citenamefont
  {Tamm}}]{Schneider.2005}%
  \BibitemOpen
  \bibfield  {author} {\bibinfo {author} {\bibfnamefont {T.}~\bibnamefont
  {Schneider}}, \bibinfo {author} {\bibfnamefont {E.}~\bibnamefont {Peik}}, \
  and\ \bibinfo {author} {\bibfnamefont {{\relax Chr}.}~\bibnamefont {Tamm}},\
  }\href {\doibase 10.1103/PhysRevLett.94.230801} {\bibfield  {journal}
  {\bibinfo  {journal} {Phys. Rev. Lett.}\ }\textbf {\bibinfo {volume} {94}},\
  \bibinfo {pages} {230801} (\bibinfo {year} {2005})}\BibitemShut {NoStop}%
\bibitem [{\citenamefont {Abdel-Hafiz}\ \emph {et~al.}(2019)\citenamefont
  {Abdel-Hafiz}, \citenamefont {Ablewski}, \citenamefont {Al-Masoudi},
  \citenamefont {Mart{\'i}nez}, \citenamefont {Balling}, \citenamefont
  {Barwood}, \citenamefont {Benkler}, \citenamefont {Bober}, \citenamefont
  {Borkowski}, \citenamefont {Bowden}, \citenamefont {Ciury{\l}o},
  \citenamefont {Cybulski}, \citenamefont {Didier}, \citenamefont
  {Dole{\v{z}}al}, \citenamefont {D{\"o}rscher}, \citenamefont {Falke},
  \citenamefont {Godun}, \citenamefont {Hamid}, \citenamefont {Hill},
  \citenamefont {Hobson}, \citenamefont {Huntemann}, \citenamefont {{Le Coq}},
  \citenamefont {{Le Targat}}, \citenamefont {Legero}, \citenamefont
  {Lindvall}, \citenamefont {Lisdat}, \citenamefont {Lodewyck}, \citenamefont
  {Margolis}, \citenamefont {Mehlst{\"a}ubler}, \citenamefont {Peik},
  \citenamefont {Pelzer}, \citenamefont {Pizzocaro}, \citenamefont {Rauf},
  \citenamefont {Rolland}, \citenamefont {Scharnhorst}, \citenamefont
  {Schioppo}, \citenamefont {Schmidt}, \citenamefont {Schwarz}, \citenamefont
  {{\c{S}}enel}, \citenamefont {Spethmann}, \citenamefont {Sterr},
  \citenamefont {Tamm}, \citenamefont {Thomsen}, \citenamefont {Vianello},\
  and\ \citenamefont {Zawada}}]{AbdelHafiz.2019}%
  \BibitemOpen
  \bibfield  {author} {\bibinfo {author} {\bibfnamefont {M.}~\bibnamefont
  {Abdel-Hafiz}}, \bibinfo {author} {\bibfnamefont {P.}~\bibnamefont
  {Ablewski}}, \bibinfo {author} {\bibfnamefont {A.}~\bibnamefont
  {Al-Masoudi}}, \bibinfo {author} {\bibfnamefont {H.~{\'A}.}\ \bibnamefont
  {Mart{\'i}nez}}, \bibinfo {author} {\bibfnamefont {P.}~\bibnamefont
  {Balling}}, \bibinfo {author} {\bibfnamefont {G.}~\bibnamefont {Barwood}},
  \bibinfo {author} {\bibfnamefont {E.}~\bibnamefont {Benkler}}, \bibinfo
  {author} {\bibfnamefont {M.}~\bibnamefont {Bober}}, \bibinfo {author}
  {\bibfnamefont {M.}~\bibnamefont {Borkowski}}, \bibinfo {author}
  {\bibfnamefont {W.}~\bibnamefont {Bowden}}, \bibinfo {author} {\bibfnamefont
  {R.}~\bibnamefont {Ciury{\l}o}}, \bibinfo {author} {\bibfnamefont
  {H.}~\bibnamefont {Cybulski}}, \bibinfo {author} {\bibfnamefont
  {A.}~\bibnamefont {Didier}}, \bibinfo {author} {\bibfnamefont
  {M.}~\bibnamefont {Dole{\v{z}}al}}, \bibinfo {author} {\bibfnamefont
  {S.}~\bibnamefont {D{\"o}rscher}}, \bibinfo {author} {\bibfnamefont
  {S.}~\bibnamefont {Falke}}, \bibinfo {author} {\bibfnamefont {R.~M.}\
  \bibnamefont {Godun}}, \bibinfo {author} {\bibfnamefont {R.}~\bibnamefont
  {Hamid}}, \bibinfo {author} {\bibfnamefont {I.~R.}\ \bibnamefont {Hill}},
  \bibinfo {author} {\bibfnamefont {R.}~\bibnamefont {Hobson}}, \bibinfo
  {author} {\bibfnamefont {N.}~\bibnamefont {Huntemann}}, \bibinfo {author}
  {\bibfnamefont {Y.}~\bibnamefont {{Le Coq}}}, \bibinfo {author}
  {\bibfnamefont {R.}~\bibnamefont {{Le Targat}}}, \bibinfo {author}
  {\bibfnamefont {T.}~\bibnamefont {Legero}}, \bibinfo {author} {\bibfnamefont
  {T.}~\bibnamefont {Lindvall}}, \bibinfo {author} {\bibfnamefont
  {C.}~\bibnamefont {Lisdat}}, \bibinfo {author} {\bibfnamefont
  {J.}~\bibnamefont {Lodewyck}}, \bibinfo {author} {\bibfnamefont {H.~S.}\
  \bibnamefont {Margolis}}, \bibinfo {author} {\bibfnamefont {T.~E.}\
  \bibnamefont {Mehlst{\"a}ubler}}, \bibinfo {author} {\bibfnamefont
  {E.}~\bibnamefont {Peik}}, \bibinfo {author} {\bibfnamefont {L.}~\bibnamefont
  {Pelzer}}, \bibinfo {author} {\bibfnamefont {M.}~\bibnamefont {Pizzocaro}},
  \bibinfo {author} {\bibfnamefont {B.}~\bibnamefont {Rauf}}, \bibinfo {author}
  {\bibfnamefont {A.}~\bibnamefont {Rolland}}, \bibinfo {author} {\bibfnamefont
  {N.}~\bibnamefont {Scharnhorst}}, \bibinfo {author} {\bibfnamefont
  {M.}~\bibnamefont {Schioppo}}, \bibinfo {author} {\bibfnamefont {P.~O.}\
  \bibnamefont {Schmidt}}, \bibinfo {author} {\bibfnamefont {R.}~\bibnamefont
  {Schwarz}}, \bibinfo {author} {\bibfnamefont {{\c{C}}.}~\bibnamefont
  {{\c{S}}enel}}, \bibinfo {author} {\bibfnamefont {N.}~\bibnamefont
  {Spethmann}}, \bibinfo {author} {\bibfnamefont {U.}~\bibnamefont {Sterr}},
  \bibinfo {author} {\bibfnamefont {{\relax Chr}.}~\bibnamefont {Tamm}},
  \bibinfo {author} {\bibfnamefont {J.~W.}\ \bibnamefont {Thomsen}}, \bibinfo
  {author} {\bibfnamefont {A.}~\bibnamefont {Vianello}}, \ and\ \bibinfo
  {author} {\bibfnamefont {M.}~\bibnamefont {Zawada}},\ }\href
  {http://arxiv.org/pdf/1906.11495v2} {\bibfield  {journal} {\bibinfo
  {journal} {arXiv:1906.11495v2}\ } (\bibinfo {year} {2019})}\BibitemShut
  {NoStop}%
\bibitem [{\citenamefont {Yu}\ and\ \citenamefont {Maleki}(2000)}]{Yu.2000}%
  \BibitemOpen
  \bibfield  {author} {\bibinfo {author} {\bibfnamefont {N.}~\bibnamefont
  {Yu}}\ and\ \bibinfo {author} {\bibfnamefont {L.}~\bibnamefont {Maleki}},\
  }\href {\doibase 10.1103/PhysRevA.61.022507} {\bibfield  {journal} {\bibinfo
  {journal} {Physical Review A}\ }\textbf {\bibinfo {volume} {61}},\ \bibinfo
  {pages} {022507} (\bibinfo {year} {2000})}\BibitemShut {NoStop}%
\bibitem [{\citenamefont {Sanner}\ \emph {et~al.}(2019)\citenamefont {Sanner},
  \citenamefont {Huntemann}, \citenamefont {Lange}, \citenamefont {Tamm},
  \citenamefont {Peik}, \citenamefont {Safronova},\ and\ \citenamefont
  {Porsev}}]{Sanner.2019}%
  \BibitemOpen
  \bibfield  {author} {\bibinfo {author} {\bibfnamefont {C.}~\bibnamefont
  {Sanner}}, \bibinfo {author} {\bibfnamefont {N.}~\bibnamefont {Huntemann}},
  \bibinfo {author} {\bibfnamefont {R.}~\bibnamefont {Lange}}, \bibinfo
  {author} {\bibfnamefont {{\relax Chr}.}~\bibnamefont {Tamm}}, \bibinfo
  {author} {\bibfnamefont {E.}~\bibnamefont {Peik}}, \bibinfo {author}
  {\bibfnamefont {M.~S.}\ \bibnamefont {Safronova}}, \ and\ \bibinfo {author}
  {\bibfnamefont {S.~G.}\ \bibnamefont {Porsev}},\ }\href {\doibase
  10.1038/s41586-019-0972-2} {\bibfield  {journal} {\bibinfo  {journal}
  {Nature}\ }\textbf {\bibinfo {volume} {567}},\ \bibinfo {pages} {204}
  (\bibinfo {year} {2019})}\BibitemShut {NoStop}%
\bibitem [{\citenamefont {Huntemann}\ \emph {et~al.}(2012)\citenamefont
  {Huntemann}, \citenamefont {Okhapkin}, \citenamefont {Lipphardt},
  \citenamefont {Weyers}, \citenamefont {Tamm},\ and\ \citenamefont
  {Peik}}]{Huntemann.2012}%
  \BibitemOpen
  \bibfield  {author} {\bibinfo {author} {\bibfnamefont {N.}~\bibnamefont
  {Huntemann}}, \bibinfo {author} {\bibfnamefont {M.}~\bibnamefont {Okhapkin}},
  \bibinfo {author} {\bibfnamefont {B.}~\bibnamefont {Lipphardt}}, \bibinfo
  {author} {\bibfnamefont {S.}~\bibnamefont {Weyers}}, \bibinfo {author}
  {\bibfnamefont {{\relax Chr}.}~\bibnamefont {Tamm}}, \ and\ \bibinfo {author}
  {\bibfnamefont {E.}~\bibnamefont {Peik}},\ }\href {\doibase
  10.1103/PhysRevLett.108.090801} {\bibfield  {journal} {\bibinfo  {journal}
  {Phys. Rev. Lett.}\ }\textbf {\bibinfo {volume} {108}},\ \bibinfo {pages}
  {090801} (\bibinfo {year} {2012})}\BibitemShut {NoStop}%
\bibitem [{\citenamefont {Porsev}\ \emph {et~al.}(2012)\citenamefont {Porsev},
  \citenamefont {Safronova},\ and\ \citenamefont {Kozlov}}]{Porsev.2012}%
  \BibitemOpen
  \bibfield  {author} {\bibinfo {author} {\bibfnamefont {S.~G.}\ \bibnamefont
  {Porsev}}, \bibinfo {author} {\bibfnamefont {M.~S.}\ \bibnamefont
  {Safronova}}, \ and\ \bibinfo {author} {\bibfnamefont {M.~G.}\ \bibnamefont
  {Kozlov}},\ }\href {\doibase 10.1103/PhysRevA.86.022504} {\bibfield
  {journal} {\bibinfo  {journal} {Phys. Rev. A}\ }\textbf {\bibinfo {volume}
  {86}},\ \bibinfo {pages} {022504} (\bibinfo {year} {2012})}\BibitemShut
  {NoStop}%
\bibitem [{\citenamefont {Nandy}\ and\ \citenamefont
  {Sahoo}(2014)}]{Nandy.2014}%
  \BibitemOpen
  \bibfield  {author} {\bibinfo {author} {\bibfnamefont {D.~K.}\ \bibnamefont
  {Nandy}}\ and\ \bibinfo {author} {\bibfnamefont {B.~K.}\ \bibnamefont
  {Sahoo}},\ }\href {\doibase 10.1103/PhysRevA.90.050503} {\bibfield  {journal}
  {\bibinfo  {journal} {Phys. Rev. A}\ }\textbf {\bibinfo {volume} {90}},\
  \bibinfo {pages} {050503(R)} (\bibinfo {year} {2014})}\BibitemShut {NoStop}%
\bibitem [{\citenamefont {Batra}\ \emph {et~al.}(2016)\citenamefont {Batra},
  \citenamefont {Sahoo},\ and\ \citenamefont {De}}]{Batra.2016}%
  \BibitemOpen
  \bibfield  {author} {\bibinfo {author} {\bibfnamefont {N.}~\bibnamefont
  {Batra}}, \bibinfo {author} {\bibfnamefont {B.~K.}\ \bibnamefont {Sahoo}}, \
  and\ \bibinfo {author} {\bibfnamefont {S.}~\bibnamefont {De}},\ }\href
  {\doibase 10.1088/1674-1056/25/11/113703} {\bibfield  {journal} {\bibinfo
  {journal} {Chinese Phys. B}\ }\textbf {\bibinfo {volume} {25}},\ \bibinfo
  {pages} {113703} (\bibinfo {year} {2016})}\BibitemShut {NoStop}%
\bibitem [{\citenamefont {Guo}\ \emph {et~al.}(2020)\citenamefont {Guo},
  \citenamefont {Yu}, \citenamefont {Liu},\ and\ \citenamefont
  {Suo}}]{Guo.2020}%
  \BibitemOpen
  \bibfield  {author} {\bibinfo {author} {\bibfnamefont {X.~T.}\ \bibnamefont
  {Guo}}, \bibinfo {author} {\bibfnamefont {Y.~M.}\ \bibnamefont {Yu}},
  \bibinfo {author} {\bibfnamefont {Y.}~\bibnamefont {Liu}}, \ and\ \bibinfo
  {author} {\bibfnamefont {B.~B.}\ \bibnamefont {Suo}},\ }\href {\doibase
  10.1088/1674-1056/ab821c} {\bibfield  {journal} {\bibinfo  {journal} {Chinese
  Phys. B}\ } (\bibinfo {year} {2020}),\ 10.1088/1674-1056/ab821c}\BibitemShut
  {NoStop}%
\bibitem [{\citenamefont {Bate}\ \emph {et~al.}(1992)\citenamefont {Bate},
  \citenamefont {Dholakia}, \citenamefont {Thompson},\ and\ \citenamefont
  {Wilson}}]{Bate.1992}%
  \BibitemOpen
  \bibfield  {author} {\bibinfo {author} {\bibfnamefont {D.~J.}\ \bibnamefont
  {Bate}}, \bibinfo {author} {\bibfnamefont {K.}~\bibnamefont {Dholakia}},
  \bibinfo {author} {\bibfnamefont {R.~C.}\ \bibnamefont {Thompson}}, \ and\
  \bibinfo {author} {\bibfnamefont {D.~C.}\ \bibnamefont {Wilson}},\ }\href
  {\doibase 10.1080/09500349214550301} {\bibfield  {journal} {\bibinfo
  {journal} {Journal of Modern Optics}\ }\textbf {\bibinfo {volume} {39}},\
  \bibinfo {pages} {305} (\bibinfo {year} {1992})}\BibitemShut {NoStop}%
\bibitem [{\citenamefont {Latha}\ \emph {et~al.}(2007)\citenamefont {Latha},
  \citenamefont {Sur}, \citenamefont {Chaudhuri}, \citenamefont {Das},\ and\
  \citenamefont {Mukherjee}}]{Latha.2007}%
  \BibitemOpen
  \bibfield  {author} {\bibinfo {author} {\bibfnamefont {K.~V.~P.}\
  \bibnamefont {Latha}}, \bibinfo {author} {\bibfnamefont {C.}~\bibnamefont
  {Sur}}, \bibinfo {author} {\bibfnamefont {R.~K.}\ \bibnamefont {Chaudhuri}},
  \bibinfo {author} {\bibfnamefont {B.~P.}\ \bibnamefont {Das}}, \ and\
  \bibinfo {author} {\bibfnamefont {D.}~\bibnamefont {Mukherjee}},\ }\href
  {\doibase 10.1103/PhysRevA.76.062508} {\bibfield  {journal} {\bibinfo
  {journal} {Phys. Rev. A}\ }\textbf {\bibinfo {volume} {76}},\ \bibinfo
  {pages} {062508} (\bibinfo {year} {2007})}\BibitemShut {NoStop}%
\end{thebibliography}
\end{document}